\documentclass[]{spie}  

\usepackage{amsmath,amsfonts,amssymb}
\usepackage{graphicx}
\usepackage[colorlinks=true, allcolors=blue]{hyperref}
\usepackage{listings}
\usepackage{orcidlink}
\usepackage{xcolor} 
\usepackage{subfig}   
 \usepackage{booktabs}

\lstdefinelanguage{TOML}{
  morecomment=[l]{\#},
  morestring=[b]",
  alsoletter={.=},
}

\lstdefinestyle{toml}{
  language=TOML,
  basicstyle=\ttfamily\footnotesize,
  columns=fullflexible,
  frame=single,
  breaklines=true,
  showstringspaces=false,
  keywordstyle=\color{blue}\bfseries,
}

\title{A Workflow-Oriented Framework for Asynchronous Human-AI Collaboration in Hybrid and Compute-Intensive HPC Environments}
\author[1,*]{Sergio Mendoza\,\orcidlink{https://orcid.org/0000-0001-8878-4310}}
\author[1]{Cedric Bhihe\,\orcidlink{0000-0002-3854-6784}}
\author[1]{Natalia Zamora\,\orcidlink{0000-0002-9754-6143}}
\author[1]{David Modesto\,\orcidlink{0000-0001-9540-8815}}
\author[2]{Jose Martin Bugallo Batalla}
\author[2]{Jesus Gomez Canovas}
\author[2]{Rafel Palomo Avellaneda}
\author[2]{Miguel Perez Espinosa}

\affil[1]{Barcelona Supercomputing Center, Spain}
\affil[2]{NTT DATA, Spain}

\authorinfo{*Correspondence: sergio.mendoza@bsc.es}

\begin{document} 
\maketitle

\noindent\begin{minipage}{\linewidth}
\hrule
\vspace{0.5em}
\textbf{Published version notice.} This is the \emph{accepted manuscript} of the paper published in the \emph{Proceedings of SPIE}.\\
DOI: \href{https://doi.org/10.1117/12.3069887}{10.1117/12.3069887}.\\
© 2025 Society of Photo-Optical Instrumentation Engineers (SPIE). One print or electronic copy may be made for personal use only. Systematic reproduction and distribution, duplication of any material in this paper for a fee or for commercial purposes, or modification of the content of the paper are prohibited.
\vspace{0.5em}
\hrule
\end{minipage}

\begin{abstract}
Human involvement is critical in training and deploying AI systems in high-stakes defence and security contexts. However, real-time interaction is impractical in HPC environments due to compute intensity and resource constraints. We present a workflow framework that enables asynchronous human-AI collaboration across hybrid infrastructures, including HPC clusters, local machines, and cloud platforms. Workflows can pause at defined checkpoints for human input without halting underlying compute jobs, preventing idle resources and enabling non-blocking supervision. The framework supports interaction with SLURM-based scheduling, containerized and native tasks, and is customized for scenarios requiring human judgment and adaptability. We demonstrate its application in model training on systems like MareNostrum 5, highlighting benefits in portability, efficiency, and oversight in operational AI workflows.
\end{abstract}


\section{Introduction}

Artificial intelligence (AI) systems are increasingly deployed in domains where reliability, adaptability, and transparency are critical, including security and defence. In these high-stakes contexts, fully automated workflows are insufficient: human expertise remains essential for oversight, validation, and adjustment. This need is addressed by the Human-in-the-Loop (HITL) paradigm, which ensures that human judgment complements algorithmic computation. When data is scarce, as in frugal learning, HITL becomes critical, since even a very small amount of expert supervision can significantly accelerate model performance\cite{jakubik2022instanceselectionmechanismshumanintheloop}.

However, traditional high-performance computing (HPC) infrastructures are not designed to allow asynchronous human participation. Once a job is submitted to a batch system such as SLURM, it is expected to execute without human interaction. This rigidity creates a misalignment between the need for human supervision and the operational characteristics of HPC. As a result, existing AI workflows cannot adapt dynamically to human input without costly interruptions, reducing both efficiency and flexibility.

Existing workflow management frameworks—such as Apache Airflow, Nextflow, or COMPSs—partially address automation, portability, and scalability, but they remain limited by code-centric definitions or static execution models. Few allow workflows to pause logically for asynchronous human supervision, and even fewer support runtime restructuring across heterogeneous infrastructures. These limits reduce their use in defence-inspired scenarios where rapid adaptation with human oversight is essential.

In response to this gap, we introduce the Collaborative Innovation Framework (CIF). CIF is both:
\begin{itemize}
    \item \textbf{A framework}: a concrete software system that enables the execution of modular workflows across HPC, cloud, and local environments.
    \item \textbf{An architecture}: a workflow-oriented architecture (WOA), where the workflow configuration file serves as the central element that defines the initial task dependencies, execution sites, and HITL checkpoints, while allowing runtime modifications that enable dynamic branching.
\end{itemize}

This dual view is essential. As a framework, CIF shows practical value through implementation and validation. As an architecture, it defines a new paradigm where configuration-based specification, asynchronous HITL checkpoints, and runtime adaptability are built-in features. By treating the workflow as the central element, CIF makes sure that human supervision and computational processes act together as complementary rather than competing forces.

The contribution of this work can be seen in two main aspects. First, it shows how workflows defined declaratively in configuration files can run across hybrid environments with integrated HITL checkpoints, allowing human supervision without stopping compute-intensive tasks. Second, it presents CIF as a workflow-oriented architecture that directly supports adaptability and rapid adaptation, ensuring that workflows can change dynamically in response to runtime conditions, such as human supervision, changing data or resource availability.

The remainder of this paper is organized as follows: Section~2 reviews related work in workflow management, hybrid execution environments, and HITL integration. Section~3 explains the design principles of CIF as a workflow-oriented architecture. Section~4 describes its implementation. Section~5 introduces a case study that validates CIF in a defence-oriented scenario. Section~6 reports the validation methodology and results. Section~7 compares CIF with existing frameworks, and Section~8 concludes with directions for future research.

\section{Related Work}

Research on workflow management and execution frameworks has advanced significantly, but important gaps remain when adaptability, asynchronous HITL checkpoints, and hybrid HPC execution need to coexist. We evaluate previous work along three main areas: workflow frameworks, hybrid execution environments, and HITL integration.

\subsection{Workflow Frameworks}

General-purpose systems such as Apache Airflow~\cite{airflow}, Luigi~\cite{luigi}, and Prefect~\cite{prefect} are widely used for task automation. They provide dependency management and monitoring, but remain code-centric (Python DAGs) and have little or no integration with HPC schedulers such as SLURM. Airflow supports branching and sensors, but these are synchronous constructs defined in Python code, not declarative configuration. This makes them harder to adapt at runtime and less portable to hybrid HPC environments.

In scientific computing, frameworks such as Nextflow~\cite{nextflow}, Snakemake~\cite{snakemake}, and Argo Workflows~\cite{argo} focus on containerization and reproducibility. These systems are effective for large-scale data pipelines but rely on fixed execution models, with no native support for asynchronous human supervision.

Closer to HPC, systems such as COMPSs~\cite{compss}, Pegasus~\cite{pegasus}, and Kepler~\cite{kepler} support large-scale distributed execution and parallelism. While powerful, their workflows remain fixed once launched and do not provide clear mechanisms to include HITL checkpoints.

\textit{Contrast with CIF:} Unlike these solutions, CIF uses a configuration-based TOML specification, includes asynchronous HITL tasks as explicit elements, and allows runtime restructuring, making it particularly suited for adaptive workflows in defence and security contexts.

\subsection{Hybrid Execution Environments}

The efforts for unifying HPC, cloud, and local resources have led to orchestration strategies such as Kubernetes Federation~\cite{k8s}, HPC-as-a-Service models~\cite{eflows4hpc_fgcs}, and scientific platforms like Galaxy~\cite{galaxy}. Community standards such as CWL~\cite{cwl} and WDL~\cite{wdl} reinforce reproducibility through declarative specifications.

Although these systems enable portability, they remain mainly machine-driven and rarely integrate human checkpoints into execution workflows. In addition, secure transfer and supervision across defence and HPC environments remain underexplored.

\textit{Contrast with CIF:} CIF directly supports hybrid execution across tactical edge resources, HPC clusters, and cloud platforms, integrating with SLURM while keeps portability through containers and explicit HITL checkpoints.

\subsection{HITL Integration in Workflows}

HITL methods are increasingly studied in medicine~\cite{holzinger_widm}, autonomous driving~\cite{ma-etal-2024-learning}, and defence operations~\cite{shneiderman2020human}. TThese approaches often rely on annotation or later validation, but interventions are synchronous, stopping the workflow until human feedback is received.

Business Process Management (BPM) systems~\cite{aalst_bpm}, as well as Airflow sensors~\cite{airflow_sensors} or n8n manual triggers~\cite{n8n_manual}, support human interaction by pausing execution until input is received. These mechanisms stop progress in their branch of execution, while other branches may continue, but they remain blocking in nature and are tied to code- or GUI-based definitions. In contrast, CIF integrates HITL checkpoints as asynchronous, declarative tasks that can change parameters or add new tasks at runtime, making them better suited for compute-intensive and hybrid HPC contexts.

\textit{Contrast with CIF:} CIF defines HITL as asynchronous, non-blocking tasks that are defined in the workflow configuration. This design supports rapid adaptation: workflows can pause logically for human decisions while parallel HPC tasks continue, ensuring resource efficiency and adaptability.

\subsection{Gap and Motivation}

Across these three areas of previous work (workflow frameworks, hybrid execution environments, and HITL integration), existing systems show two main patterns: (i) they support automation without human adaptability, or (ii) they include human oversight in ways that are not compatible with HPC. None combine configuration-based workflows, asynchronous HITL, runtime restructuring, and hybrid HPC execution in a single approach.

The CIF addresses this gap by enabling dynamic workflows with HITL, ensuring adaptability and rapid adaptation in hybrid, compute-intensive environments. Current workflow systems deal with automation, hybrid execution, or human interaction only on their own, but none provide a unified solution that brings these aspects together. Many are optimized for deterministic automation without human adaptability, while others manage human oversight in ways that do not work well in HPC environments. This motivates the design of CIF, which defines asynchronous HITL checkpoints as workflow tasks, supports dynamic restructuring at runtime, and keeps execution running across different infrastructures. 

The following section introduces the architectural principles of CIF, showing how configuration-based specification, runtime orchestration, and hybrid execution are integrated into a workflow-oriented architecture designed for adaptability and rapid adaptation under human-AI collaboration~\cite{gomez2025taxonomy,hemmer2024complementarity}.

\section{Framework Architecture and Design}

\subsection{Workflow-Oriented Architecture (WOA)}

The CIF adopts a workflow-oriented architecture (WOA) in which the workflow configuration is not only an execution input but the element that organizes the system. Workflows are written declaratively in TOML files, defining tasks, dependencies, execution sites, and HITL checkpoints. At runtime, the \textit{TaskScheduler} reads the configuration, activates tasks once dependencies are satisfied, and submits them to the right execution environment. By separating workflow logic from implementation code, CIF reduces barriers for non-developer users and improves portability.

A key property of this architecture is that HITL checkpoints are defined in the same configuration file as computational tasks, making supervision an integral part of workflow logic. In addition, workflows can be changed during execution: tasks can be added, parameters revised, or branches redirected based on human decisions. In short, WOA assures that CIF workflows remain adaptive and portable, while treating HITL checkpoints as structural rather than optional elements.

\subsection{Workflow-Oriented System (WOS)}

The architecture is implemented through lightweight modules that together form a workflow-oriented system (WOS). At the entry point, CIF-CLI checks the TOML workflow file and submits it to the CIF core components. The \textit{TaskScheduler} then interprets the configuration, resolves dependencies, coordinates task execution, and manages pauses at HITL checkpoints and branching. The \textit{TaskExecutor} handles input/output transfers, execution environments, and interactions with local, cloud, or HPC schedulers such as SLURM. The \textit{WorkflowExecution} component maintains the global workflow state. The \textit{Watcher} checks completions asynchronously in HPC environments by polling for \texttt{.done} markers, while the \textit{JobRegistry} tracks status of all tasks to ensure consistency across computational and HITL tasks.

This modular design allows CIF to integrate external frameworks (e.g., TensorFlow, MPI, COMPSs) without modifying the workflow definition, while ensuring robust state management in hybrid and asynchronous contexts.

\begin{figure}[ht]
\centering
\includegraphics[width=0.85\linewidth]{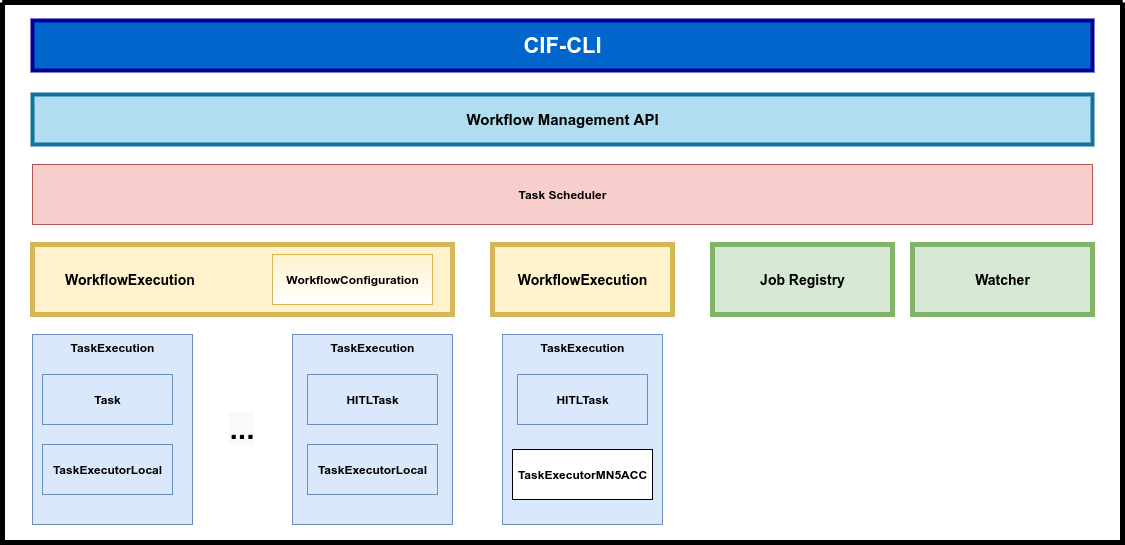}
\caption{Layered architecture of CIF. At the top, CIF-CLI receives the workflow file and starts execution. The middle layer (TaskScheduler, TaskExecutor, WorkflowExecution) organizes tasks and controls their flow. The Watcher and JobRegistry manage monitoring and state across the system. At the bottom, tasks run on different sites (local, HPC, cloud). This layered view shows that all modules act according to the workflow, making CIF workflow-centric, adaptive, and ready for asynchronous HITL.}
\label{fig:cif_architecture}
\end{figure}

Together, these modules formalize CIF’s capacity for adaptability, hybrid execution, and workflow management with HITL integration.

\subsection{Workflow Execution Model}

CIF workflows follow a dependency-driven execution model: the \textit{TaskScheduler} starts tasks when their prerequisites are complete, the \textit{TaskExecutor} runs them, and \textit{WorkflowExecution} keeps the global state updated. Unlike static directed acyclic graphs (DAGs), CIF supports runtime restructuring. HITL checkpoints can add new tasks or change execution paths while computation continues in parallel branches. In Figure~2, you can see that Task~A runs locally while Task~B is running in MareNostrum~5 (MN5)\footnotemark[2], showing how the same workflow can run across different execution sites. This model makes sure that HITL steps do not block ongoing HPC jobs. While one branch may wait for human feedback, others continue normally, keeping efficiency. In this way, CIF enables rapid adaptation by making sure that waiting for human feedback does not slow down the computing tasks.

\footnotetext[2]{https://www.bsc.es/supportkc/docs/MareNostrum5/intro/}

\subsection{Hybrid Execution and Containerization}

Workflows may extend across tactical edge resources\textit{tactical cloudlets} (deployable computing nodes operating near the mission area), HPC clusters, cloud platforms, and local machines. Execution sites are declared directly in the TOML file, allowing the same workflow to distribute tasks across heterogeneous infrastructures. Containerization is an explicit feature: Docker and Singularity images preserve dependencies and runtime consistency across environments. In CIF, each task can reference a specific container image in the workflow file. When the \textit{TaskExecutor} runs the task, it uses the defined container to ensure the task always runs in the same environment. In this way, CIF avoids inconsistencies between environments and makes it easier to integrate external frameworks. This hybrid strategy guarantees portability and flexibility while supporting adaptability in mission-oriented contexts.

\subsection{Asynchronous HITL Integration}

HITL checkpoints are defined as declarative tasks within the workflow configuration. When reached, the workflow pauses logically for human review (for example, validating inference or authorizing retraining) without stopping parallel jobs already running on HPC or cloud resources. The \textit{Watcher} and \textit{JobRegistry} manage the asynchronous use of human decisions, which may adjust parameters, trigger new tasks, or skip steps that are no longer needed. This guarantees supervision without leaving resources idle.

\begin{figure}[ht]
\centering
\includegraphics[width=0.85\linewidth]{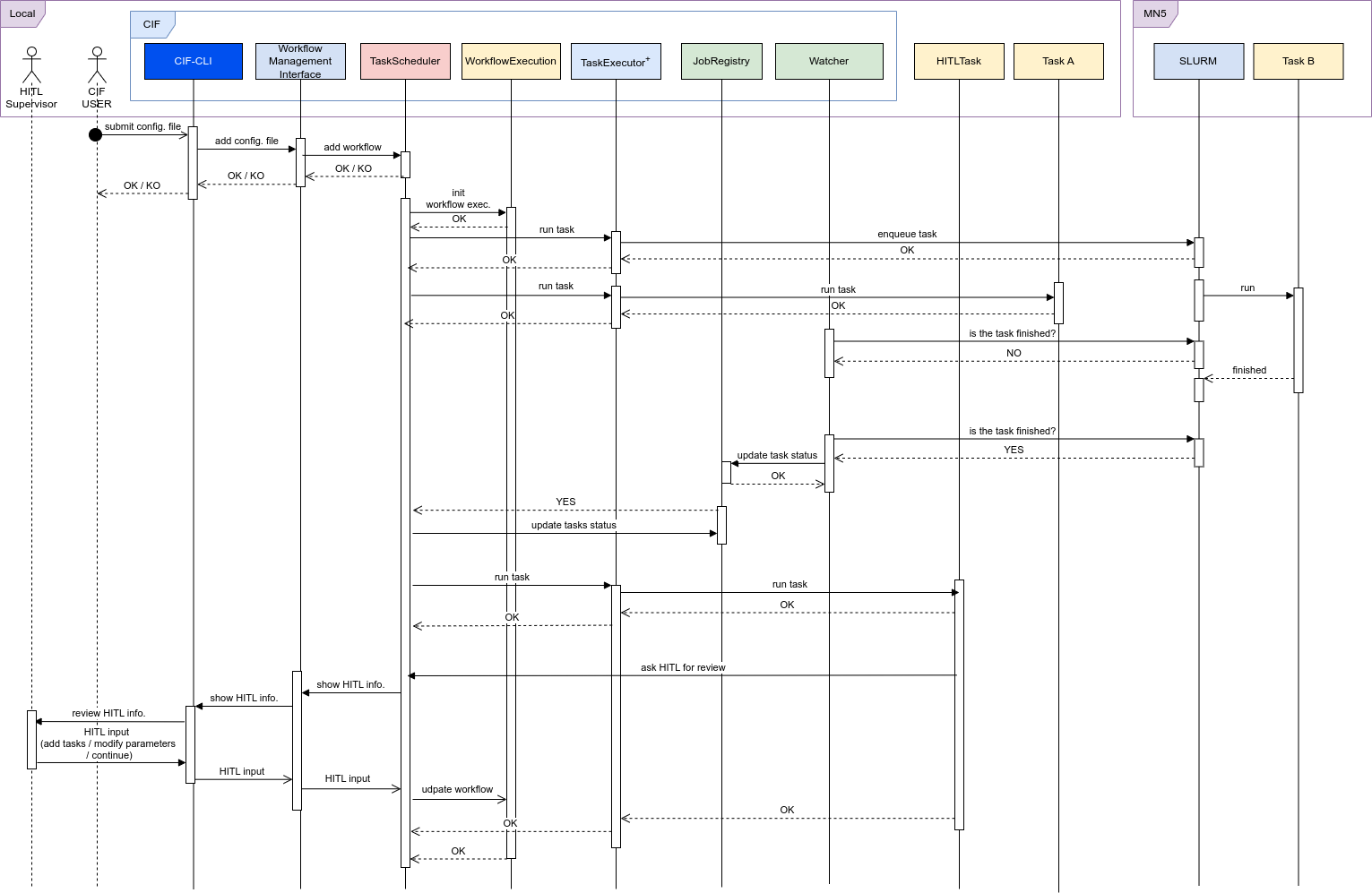}
\caption{UML sequence of asynchronous HITL in CIF. Two stakeholders interact through the CIF-CLI: the CIF user (who submits the workflow) and the HITL supervisor (who reviews and adapts it). The user submits a configuration file in TOML, which is parsed by the TaskScheduler. Tasks are executed on different sites, such as MN5 via SLURM, through dedicated TaskExecutors. The Watcher polls HPC jobs and notifies the JobRegistry when tasks finish. At HITL checkpoints, the supervisor provides feedback through CIF-CLI, such as adding tasks or changing parameters. The TaskScheduler then updates the workflow and continues execution. This sequence shows how CIF keeps workflows running while separating human review time from compute time.}
\label{fig:hitl_sequence}
\end{figure}

By defining HITL as asynchronous, non-blocking tasks, CIF balances adaptability and efficiency, which is critical in defence and security operations.

\subsection{Integrated State Management and Differentiation}

In HPC contexts, schedulers such as SLURM provide no workflow-level notifications. CIF addresses this with the \textit{Watcher}, which detects completion signals, and the \textit{JobRegistry}, which keeps the central record of workflow state. Together, these components ensure integrity across asynchronous tasks.

Architecturally, CIF distinguishes itself by combining three principles: first, a workflow-oriented architecture, where workflows act as the structural axis rather than just execution inputs; second, asynchronous HITL checkpoints, which are non-blocking and declared in the workflow; and third, hybrid execution designed for HPC, achieved through direct integration with SLURM and containerization technologies. In short, CIF establishes a new model where configuration-based workflows can adapt dynamically under human supervision while keeping full use of compute resources.

\section{Implementation}

This section describes how the CIF is implemented in practice. Building on the workflow-oriented architecture introduced in Section~3, the implementation shows how configuration-based workflow control, hybrid execution, and asynchronous HITL supervision are applied in workflows.

\subsection{CIF-CLI and Workflow Submission}

The CIF command-line interface (CIF-CLI) is the entry point for workflow execution. 
It serves as the user interface for submitting workflows to the CIF framework. 
The workflow is described in a configuration file written in TOML, which specifies tasks, 
dependencies, and execution sites. Once submitted through CIF-CLI, the file is interpreted 
by the \textit{TaskScheduler}, which parses the configuration and manages the execution 
flow by starting tasks when dependencies are satisfied and sending them to the correct 
execution environment.

\begin{lstlisting}[language=toml, caption={Minimal CIF configuration file: task and workflow (TOML).}, label={code:minimal_task}]
[[task]]
name = "inference_task"
command = "singularity run --nv \
--bind <DATA_ROOT> <SIF_IMAGE> \
src/main.py --config <CONFIG_FILE>"
execsite = "local"
input.model = "/path/to/model.pt"
input.data = "/path/to/images/"
output.results = "/path/to/results.json"

[workflow]
tasks = [
    "inference_task"
]
\end{lstlisting}

Code~\ref{code:minimal_task} shows the simplest configuration of a CIF workflow. 
It has two sections: first, the task definition, and second, the workflow description. 
The single task is named \texttt{inference\_task}. Its command is a Singularity 
execution line written in Bash, which will be launched in the operating system shell. 
The field \texttt{execsite = "local"} indicates that the task runs in the same machine 
where CIF is executed. The task requires two input paths: a model file (\texttt{model.pt}) 
and a dataset directory. As output, it produces a JSON file with the inference results. 
The \texttt{[workflow]} section lists this task for execution.

This design keeps the workflow description in a simple configuration file, while CIF modules 
(\textit{TaskScheduler}, \textit{TaskExecutor}) take care of parsing and running the tasks. 
As a result, workflows are easy to read, portable across environments, and can be adapted 
at runtime without changing the underlying application code.

\subsection{Multi-Stage HPC Workflow}

CIF supports multi-stage workflows on HPC systems by declaring dependencies between tasks such as training and evaluation. A multi-stage workflow is one where tasks are chained in order, with each stage depending on the output of the previous one. These tasks are run in the execsite specified in the TOML file. Unlike code-centric DAGs, CIF workflows remain declarative and portable.

For example, a training task may generate a model, followed by an evaluation task that depends on it. Both are scheduled on the HPC cluster. While the detailed TOML definition is omitted here, the setup follows Code~\ref{code:minimal_task}, with explicit dependencies and execution sites declared. This demonstrates CIF’s ability to scale workflows declaratively on HPC without changing the configuration model.

\subsection{HITL Checkpoint Integration}

A key feature of CIF is the ability to include Human-in-the-Loop (HITL) checkpoints directly as tasks in the workflow configuration file. These checkpoints allow a human supervisor to review metrics or outputs and decide how the workflow should continue. Unlike blocking mechanisms in other systems, CIF HITL tasks are asynchronous: they pause only their workflow branch, while unrelated tasks continue executing in parallel. This behaviour is consistent with the execution model shown in Figure~\ref{fig:hitl_sequence}, where \textit{Task~A} and \textit{Task~B} run in parallel on different execution sites.

\begin{lstlisting}[language=toml, breaklines=true, caption={Declarative definition of an asynchronous HITL checkpoint task}, label={code:hitl_task}]
[[task]]
name = "training_task"
command = "singularity run --nv \
--bind <DATA_ROOT> <SIF_IMAGE> \
src/main.py --config <CONFIG_FILE>"
depends_on = ["inference_task"]
execsite = "HPC-CLUSTER"
input.config = "<CONFIG_FILE>"
output.metrics = "<OUTPUT_MODELS_DIR>"

[[task]]
name = "hitl_reviewer_evaluation"
command = "modules.cif_core.hitl_review.hitl_epoch_reviewer"
execsite = "HPC-CLUSTER"
depends_on = ["training_task"]
add_tasks = ["inference_task"]

hitl.enabled = true
hitl.input = "/path/to/output/metrics.txt"
hitl.message = "Is the model good enough?\n"
output.hitl_decision = "hitl_decision.json"

[execsites."HPC-CLUSTER"]
host = "hpc-login.example.org"
key = "/home/user/.ssh/id_ecdsa"
user = "user1234"

[workflow]
tasks = [ "training_task", "hitl_reviewer_evaluation"]
\end{lstlisting}

Code~\ref{code:hitl_task} shows how CIF defines an asynchronous HITL checkpoint. 
Two tasks are included. The first, \texttt{training\_task}, depends on the inference stage and is executed on the HPC cluster. Its command is a Singularity execution line in Bash that trains the model, using a configuration file as input and producing metrics as output. 

The second task, \texttt{hitl\_reviewer\_evaluation}, also runs on the HPC cluster and depends on the training task. Instead of a compute-intensive operation, it calls the CIF HITL review module, which presents metrics to the human supervisor. This task is marked with \texttt{hitl.enabled = true}, meaning that only this workflow branch pauses until the supervisor 
provides input. Other tasks without this dependency continue to run. The supervisor’s decision is saved in a JSON file (\texttt{hitl\_decision.json}) and may include adding new tasks through the \texttt{add\_tasks} field. 

The section \texttt{[execsites."HPC-CLUSTER"]} defines the CIF needs for using to the HPC system, showing that CIF itself runs externally and only submits tasks remotely. The workflow section at the bottom lists both tasks, confirming that HITL checkpoints are treated as regular workflow tasks. 

This example also shows that dependencies in CIF are not mandatory. Tasks can run independently, in parallel, or as chained stages. Workflows are not limited to linear pipelines, since HITL checkpoints allow branching and the dynamic addition of tasks. Loops could in principle be described directly in TOML by declaring cyclic dependencies, but this would be unsafe in HPC environments. For this reason, CIF discourages explicit loops and instead supports iterative processes through HITL-driven task addition. This mechanism enables retraining cycles while keeping execution predictable and safe.

Overall, HITL checkpoints in CIF are integrated as asynchronous, non-blocking elements. They allow supervisors to influence workflow paths dynamically while other branches continue, supporting adaptability and rapid adaptation without stopping ongoing computation.

\subsection{Execution Flow Example}

The CIF execution cycle can be summarized as:

\begin{center}
\texttt{Parse $\rightarrow$ Schedule $\rightarrow$ Execute $\rightarrow$ HITL Review $\rightarrow$ Adapt $\rightarrow$ Continue}
\end{center}

This flow is illustrated in Figure~\ref{fig:hitl_sequence}. It shows how a workflow is first parsed, tasks are scheduled, execution starts, and then HITL checkpoints may pause the flow for review. Human input can adapt the workflow, after which execution continues.

Unlike systems such as Airflow~\cite{airflow_sensors} or n8n~\cite{n8n_manual}, where workflow control is either code-centric or limited to blocking steps, CIF defines HITL and hybrid execution as declarative features. This design combines portability with HPC-native execution and allows workflows to adapt at runtime by adding new tasks or changing parameters. As a result, CIF workflows remain dynamic, adaptive, and quick to reconfigure, which is especially important in mission-oriented HPC scenarios.

\section{Case Study: Tactical Use Case Execution}

\subsection{Description of the Use Case}

The tactical case study focuses on ship detection in maritime surveillance, where operational needs require both real-time inference at the edge and adaptive retraining on HPC resources. A Tactical Unit deployed in the field runs inference tasks on image streams, while supervisors review outputs asynchronously through HITL checkpoints. If inference performance is not good enough, CIF manages retraining and evaluation on the MN5 supercomputer.

\begin{figure}[ht]
\centering
\includegraphics[width=0.85\linewidth]{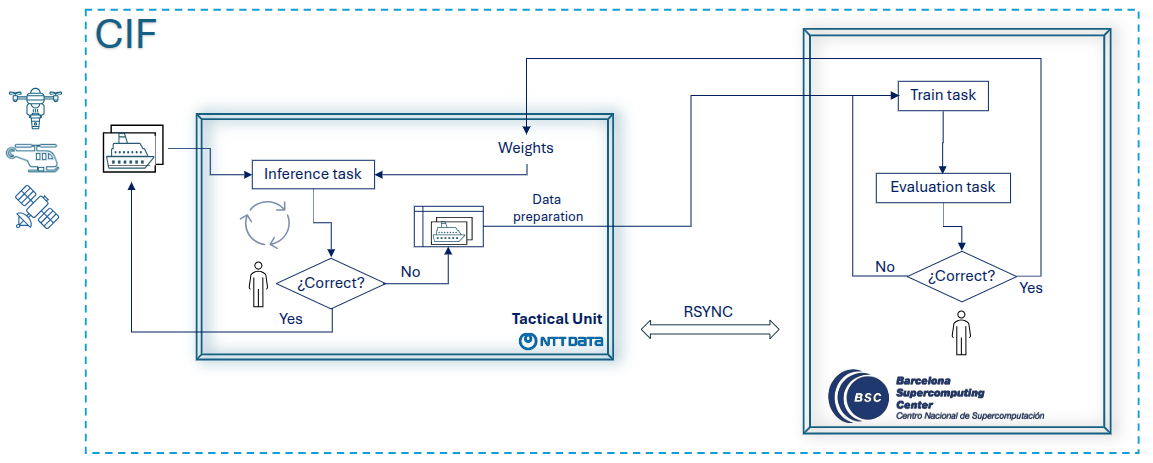}
\caption{Abstract workflow for the ship-detection case study. Two execution sites are used: the Tactical Unit and MN5 --- HPC Resources in BSC. In a real mission, input images would come from drones, satellites, or other sensors (left side). Inference runs on the Tactical Unit because it is lightweight, while training is compute-intensive and runs on MN5. Two HITL checkpoints are included: one where the supervisor reviews inference results and decides whether to trigger retraining, and another where the updated model is validated after training to decide if it can be deployed or needs improvement.}
\label{fig:case_study_overview}
\end{figure}

The overall workflow is illustrated in Figure~\ref{fig:case_study_overview}, which shows how CIF links local inference, human supervision, and compute-intensive retraining across different environments.

\subsection{Workflow Structure and Execution Plan}

The workflow is written declaratively in a TOML configuration file and includes three categories of tasks: inference tasks that run locally on the Tactical Unit in Singularity containers; HITL checkpoints where supervisors validate detections asynchronously; and retraining and evaluation tasks that run on MN5, also in Singularity containers under SLURM.

\begin{figure}[ht]
\centering
\includegraphics[width=0.95\linewidth]{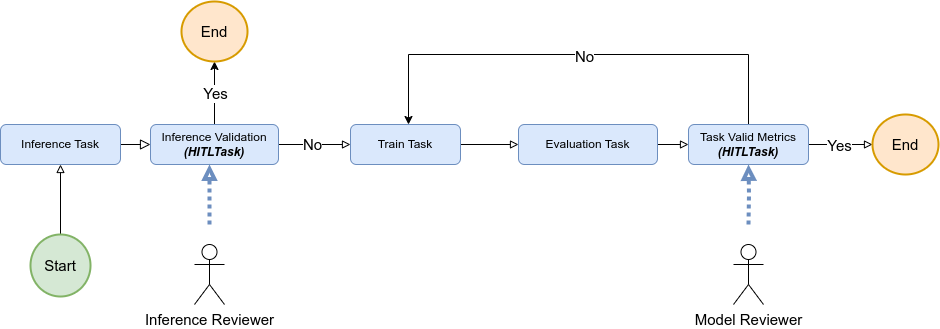}
\caption{Workflow structure for the ship-detection case study. The figure shows the initial configuration of the workflow defined in the configuration file. Dependencies between tasks are visible together with two HITL checkpoints. The Inference Reviewer can approve results or trigger retraining on MN5, while the Model Reviewer validates the retrained model. This structure illustrates how CIF workflows adapt dynamically under human supervision.}
\label{fig:case_study_workflow}
\end{figure}

As shown in Figure~\ref{fig:case_study_workflow}, inference is followed by a HITL checkpoint (Inference Validation). If the results are accepted, the workflow ends. If they are rejected, CIF runs retraining and evaluation on MN5. A second HITL checkpoint validates the metrics before deployment. In short, CIF makes sure that the workflow can change dynamically under human oversight, combining local/edge inference with HPC-scale retraining.

\subsection{Role of HITL in Supervision and Rapid Adaptation}

HITL supervision was essential for adaptability. At runtime, supervisors could reject detections or request retraining directly through HITL checkpoints, without stopping other workflow branches. This made it possible to add new tasks, such as retraining jobs, into the workflow while it was running. This mechanism shows rapid adaptation: the workflow changes in response to human feedback while compute-intensive retraining continues asynchronously on MN5. Instead of re-submitting jobs or modifying code, supervisors influenced execution by providing decisions that updated the workflow state. In this way, CIF showed that asynchronous HITL checkpoints let mission workflows adapt in real time without reducing computational efficiency.

\subsection{Integration with CIF and Containerization}

CIF managed the case study workflow through declarative specifications, using the same TOML file across the Tactical Unit and MN5. Workflow logic was kept separate from infrastructure, which ensured portability between local and HPC execution. Containerization provided consistent environments: on the Tactical Unit, Singularity was used for the Inference, Inference Validation, and Task Valid Metrics tasks. On MN5, Singularity was also used for the Evaluation Task when retraining the model. Native Python modules were also supported for evaluation steps, showing flexibility.

Operational limits, such as CPU-only inference on the Tactical Unit and restricted network bandwidth, were reduced by CIF’s hybrid design. Inference was run locally on the Tactical Unit, while only model updates or evaluation results were sent to MN5. This avoided heavy data transfers and kept response times low. In contrast, training the model required many GPUs and was too compute-intensive to be executed on the Tactical Unit, so it was performed on MN5. The framework also supported effective cooperation between the Tactical Unit, HPC resources, and asynchronous human supervision. Feedback from NTTDATA confirmed that HITL checkpoints were essential for supervisors’ confidence in the detection results, and that defining tasks in CIF was simple and required little extra training.

\section{Validation and Demonstration}

This section evaluates whether the CIF can run hybrid workflows with asynchronous HITL checkpoints while keeping computational efficiency. Building on the architectural principles in Section~3, the implementation in Section~4, and the case study in Section~5, we test whether CIF keeps execution continuous across heterogeneous resources and enables rapid adaptation to human feedback under operational constraints. We focus on three aspects: workflow continuity with HITL, resource efficiency during hybrid execution, and adaptability through runtime changes.

\subsection{Objective of the Validation}

The validation aimed to confirm that CIF supports dynamic adaptation to human feedback without disrupting ongoing HPC workloads. In particular, we examined whether asynchronous HITL checkpoints could change execution paths, including triggering retraining, while parallel tasks remained active. We also tested whether this mechanism enabled timely model improvement in a mission-oriented setting. Success is defined as achieving human-guided reconfiguration of the workflow without leaving compute resources idle and without modifying code in running components.

\subsection{Methodology of Demonstration}

\begin{figure}[ht]
\centering
\includegraphics[width=0.65\linewidth]{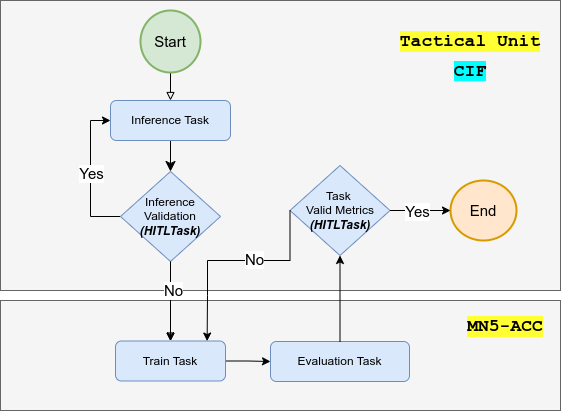}
\caption{Validation workflow for CIF (flowchart view). Inference runs on the Tactical Unit, while training and evaluation are executed on MN5-ACC. The CIF framework and CIF-CLI operate on the Tactical Unit and control remote tasks without being installed on MN5. Two HITL checkpoints guide the flow: inference validation and model validation. This figure shows how CIF combines local inference on the Tactical Unit, remote HPC training, and asynchronous HITL to adapt workflows dynamically.}
\label{fig:validation_workflow}
\end{figure}

Validation was conducted on a hybrid setup combining a field-deployed Tactical Unit for on-device inference and the MN5 supercomputer for retraining and evaluation. As shown in Figure~\ref{fig:validation_workflow}, the workflow proceeded as follows: the Tactical Unit executed local inference on incoming imagery, after which an asynchronous HITL checkpoint allowed a supervisor to validate detections. If detections were rejected by the supervisor, the CIF automatically scheduled retraining and evaluation tasks on MN5. Once these tasks finished, a second HITL checkpoint determined whether the updated model was approved for deployment.

Throughout, CIF’s \textit{Watcher} polled for completion signals, and the \textit{JobRegistry} tracked the status of running tasks. This information allowed WorkflowExecution to apply HITL decisions while keeping the workflow consistent and avoiding conflicts between human input and ongoing computation.

This setup showed that human decisions were included in the workflow while HPC jobs kept running at full capacity. The workflow paused only in a logical way at HITL checkpoints, without stopping other tasks. This means that while the system waited for human feedback, other tasks could continue running on the supercomputer. In this way, CIF kept human review in the loop without wasting compute resources.

\subsection{Adaptability through Iterative Retraining}

A key feature of CIF is its ability to let workflows be changed during execution through HITL input. HITL checkpoints allowed supervisors to reject outputs and start new retraining cycles without re-submitting jobs or changing source code. The initial TOML configuration listed the possible tasks and checkpoints, and the \textit{TaskScheduler} chose the path based on human input saved in decision files. This design made rapid adaptation possible: when supervisors identified low confidence in detections or changes in the data (e.g. new types of ships, difficult weather conditions or suspicious camouflage), they could modify the workflow, while other computations continued on MN5 or on the Tactical Unit.

Such adaptability is important in mission contexts where data or adversarial behavior may change unpredictably. By including HITL as an explicit, non-blocking element, CIF enabled supervisors to keep the system responsive without reducing throughput or wasting resources, supporting retraining when needed and avoiding unnecessary cycles when it was not.

\subsection{Results}

CIF successfully coordinated inference on the Tactical Unit with retraining and evaluation on MN5, while keeping compute resources fully used. Human supervision guided the execution path at two points: initial inference validation and approval after retraining. When the first checkpoint showed poor metrics, the HITL told the CIF to run retraining and evaluation tasks without blocking parallel activity. When the metrics met operational thresholds, the updated model was approved for deployment, closing the adaptation loop.

\begin{figure}[ht]
\centering
\subfloat[Baseline inference\label{fig:6_4a}]{
  \includegraphics[width=0.48\linewidth]{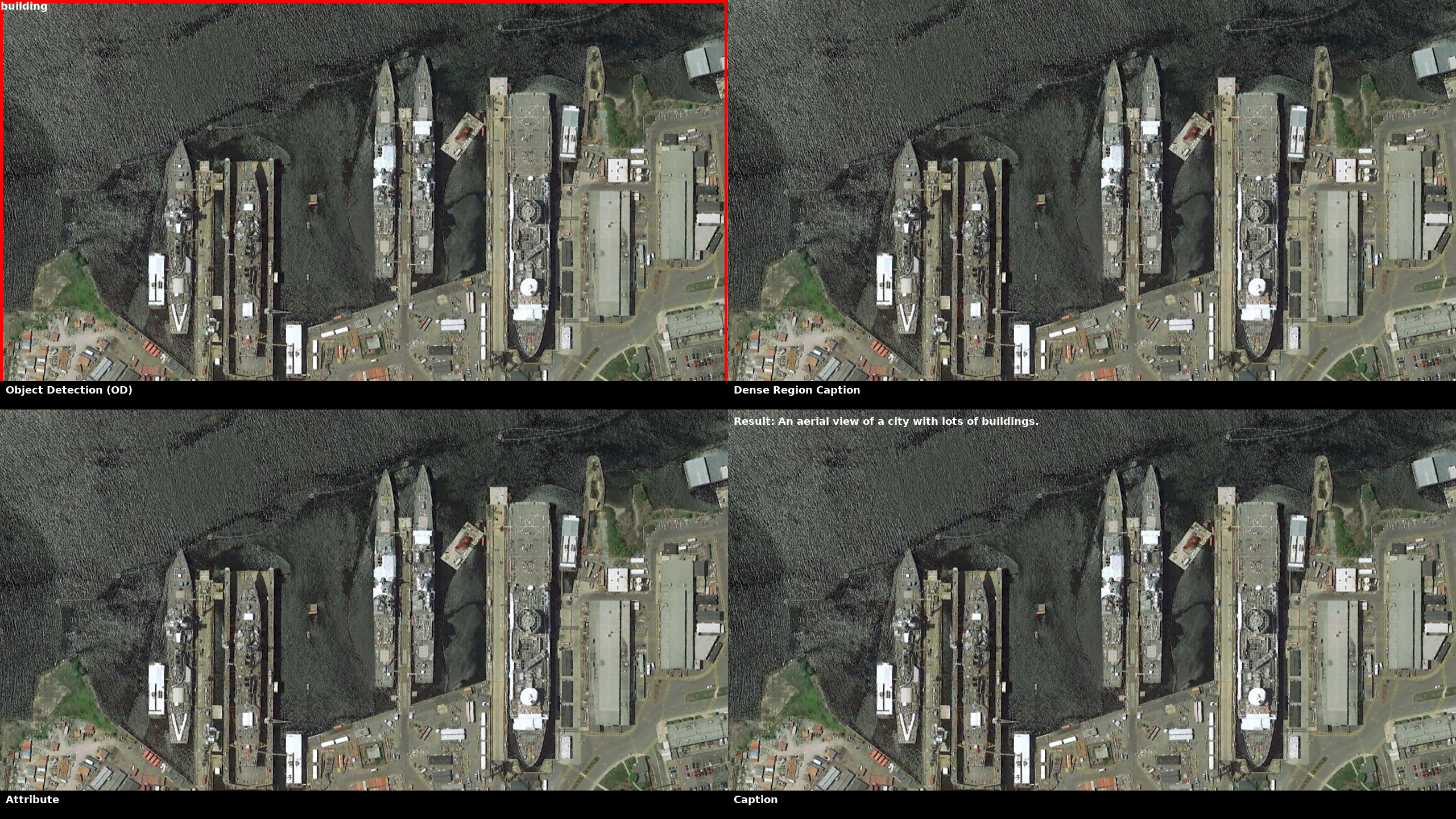}
}
\hfill
\subfloat[Post-retraining inference\label{fig:6_4b}]{
  \includegraphics[width=0.48\linewidth]{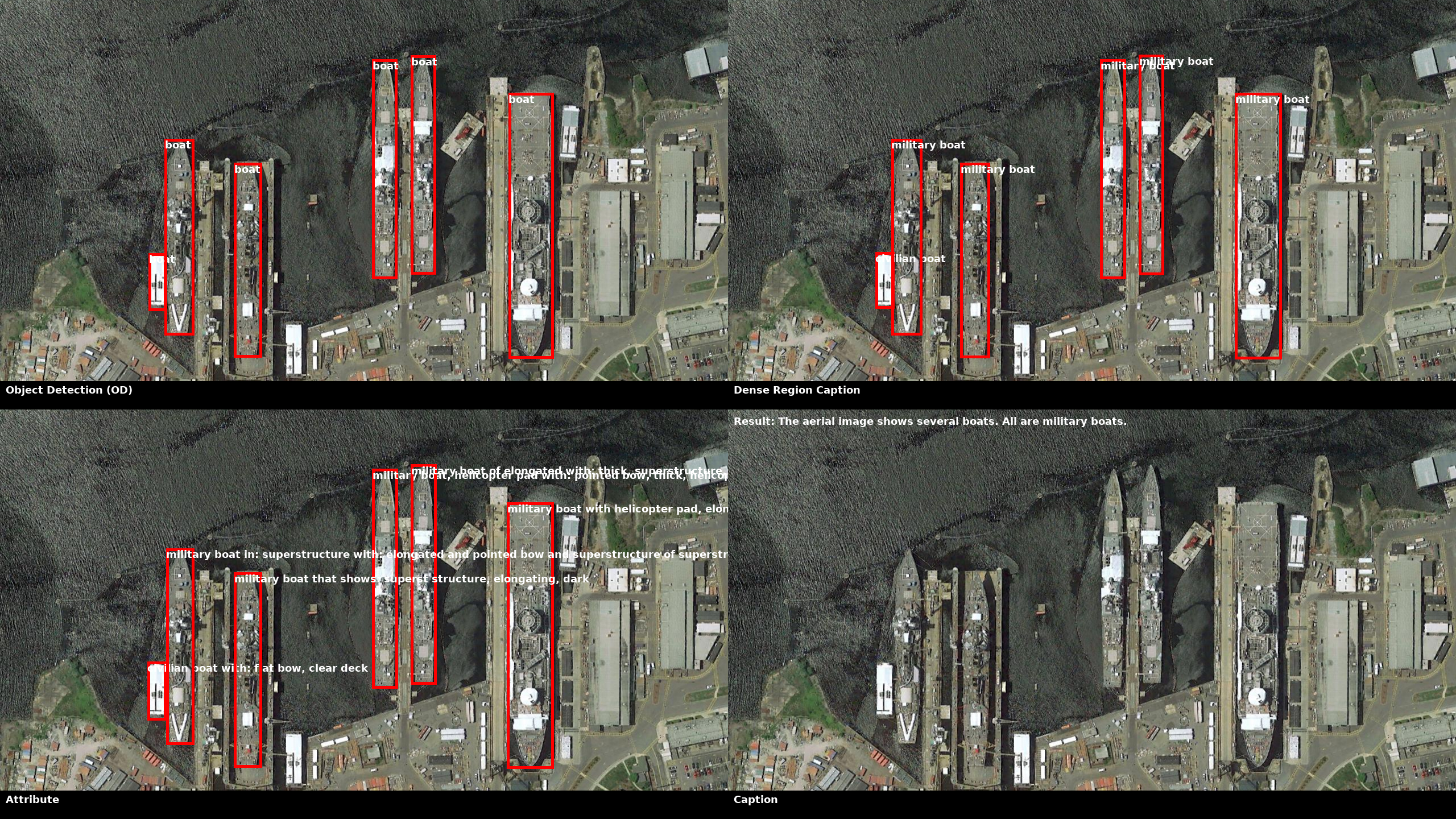}
}
\caption{Qualitative inference results before and after HITL-supervised retraining. (a) Baseline: generic ``vehicle'' detection, missing ships. (b) Post-retraining: multiple ships detected with higher precision.}
\label{fig:6_4}
\end{figure}

As shown in Figure~\ref{fig:6_4a}, the baseline model produced a generic ``building'' detection and missed ships of interest. After the HITL triggered the retraining task on MN5, Figure~\ref{fig:6_4b} shows that multiple ships were correctly detected with higher precision. These qualitative results confirm that CIF executed the adaptive loop as intended: HITL checkpoints identified errors, CIF scheduled retraining, the model was updated, and redeployment followed without interruption. Although precise model optimization was not the primary goal of this study, the improved detections illustrate CIF’s ability to support dynamic, human-guided adaptation while keeping computational efficiency. Overall, the demonstration validates that configuration-based workflows with asynchronous HITL can sustain execution continuity across heterogeneous infrastructures while enabling rapid, mission-relevant updates.

\section{Discussion: Comparison with Existing Frameworks and Limitations}

This section contrasts the CIF with representative workflow systems used in data engineering and scientific computing, and then discusses limitations and applicability. Our analysis focuses on five dimensions that are central to this work: (i) configuration-based workflow specification, (ii) hybrid/HPC-native execution, (iii) containerization for reproducibility, (iv) asynchronous HITL checkpoints, and (v) runtime restructuring guided by human supervision. We select widely adopted workflow managers—Airflow~\cite{airflow}, Luigi~\cite{luigi}, Prefect~\cite{prefect}, Nextflow~\cite{nextflow}, Snakemake~\cite{snakemake}, Argo~\cite{argo}, COMPSs~\cite{compss}, Pegasus~\cite{pegasus}, and Kepler~\cite{kepler}—as baselines for comparison.

\subsection{Comparative Analysis}

This comparison builds on the workflow systems introduced in Section~2, focusing on five aspects: configuration-based specification, hybrid execution, containerization, asynchronous HITL, and runtime adaptation. Table~\ref{tab:framework_comparison} shows the main differences.

\begin{table}[ht]
\centering
\caption{Comparison of selected workflow frameworks against CIF.}
\label{tab:framework_comparison}
\resizebox{\textwidth}{!}{%
\begin{tabular}{lcccccc}
\toprule
\textbf{Framework} & \textbf{Workflow spec.} & \textbf{HPC/Hybrid} & \textbf{Containers} & \textbf{Async HITL} & \textbf{Runtime restruct.} & \textbf{Scheduler integ.} \\
\midrule
Airflow~\cite{airflow}     & Code (Python DAG) & Limited\footnotemark[1] & Partial & No  & Limited (manual) & Plugins / ext. \\
Luigi~\cite{luigi}         & Code (Python)     & Limited                 & Partial & No  & Limited          & Plugins / ext. \\
Prefect~\cite{prefect}     & Code (Python)     & Limited                 & Partial & No  & Limited          & Plugins / ext. \\
Nextflow~\cite{nextflow}   & DSL / Config      & Yes (Cloud/HPC)         & Yes     & No  & Rare             & Native / adapters \\
Snakemake~\cite{snakemake} & Rules (Py/Config) & Yes (HPC)               & Yes     & No  & Rare             & Native / adapters \\
Argo~\cite{argo}           & YAML (K8s)        & Cloud/K8s               & Yes     & No  & Limited          & K8s-native \\
COMPSs~\cite{compss}       & Annotations/Conf. & Yes (HPC)               & Partial & No  & No               & Native (HPC) \\
Pegasus~\cite{pegasus}     & DAX/Config        & Yes (HPC/Grid)          & Partial & No  & No               & Native / schedulers \\
Kepler~\cite{kepler}       & Config/GUI        & Yes (HPC)               & Partial & No  & No               & Native / schedulers \\
\midrule
\textbf{CIF (this work)}   & \textbf{Config (TOML)} & \textbf{Yes (Edge/HPC/Cloud)} & \textbf{Yes} & \textbf{Yes (non-blocking)} & \textbf{Yes (HITL-driven)} & \textbf{SLURM + ext.} \\
\bottomrule
\end{tabular}
}
\end{table}

General-purpose systems (Airflow, Luigi, Prefect) provide dependency management and monitoring, but they are code-centric and link to HPC only through extensions. Scientific workflow managers (Nextflow, Snakemake, Argo) focus on portability and containerization, making them effective for large-scale pipelines. However, human oversight in these systems is usually synchronous and blocking or handled outside the workflow. They may allow manual approval steps, but they do not support runtime updates where human input changes the workflow itself. HPC-focused frameworks (COMPSs, Pegasus, Kepler) enable parallel execution and traceability, but workflows remain static after submission and do not include asynchronous HITL as an explicit feature.

CIF differs in three ways. First, the workflow configuration file (TOML) is the central element of execution. Second, asynchronous HITL checkpoints are defined directly as workflow tasks, allowing human supervision without blocking other branches. Third, execution sites are declared explicitly, so the same configuration can run locally, on tactical units, HPC clusters under SLURM or cloud resources. The \textit{TaskScheduler}, \textit{Watcher}, and \textit{JobRegistry} support runtime changes based on HITL decisions while keeping compute resources fully used. In this way, CIF separates the time needed for human review from the time needed for computation, which is essential in mission contexts where rapid adaptation is required.

\footnotetext[1]{Hybrid execution in these systems is usually achieved by relying on external executors, instead of providing direct integration with HPC environments.}

From an operational viewpoint, CIF’s main contribution is making HITL both \emph{declarative} and \emph{asynchronous}. Instead of stopping the whole workflow while waiting for approval, CIF lets human validation happen in parallel with other compute tasks. This design keeps HPC resources productive and enables rapid model updates when supervision detects errors, as shown in the ship-detection case study in Section~6. The configuration-based approach also makes CIF easier to adopt. Users can adjust execution by editing configuration or decision files, without having to modify the application code.

\subsection{Limitations and Applicability}

Users of CIF must be aware of several constraints: (i) each execution site must support containers (Docker or Singularity); (ii) HPC systems must provide a batch scheduler (e.g., SLURM) with accessible login nodes; (iii) task completion is detected through file-based signals (\texttt{.done} markers) used by the \textit{Watcher}, with custom extensions needed at sites that do not allow containerization or external polling; and (iv) users must have the necessary accounts and access rights to submit jobs on remote resources (e.g. MN5 or equivalent HPC systems).

In addition to these technical constraints, the duration of the adaptive loop depends on external factors such as data transfer times and queue delays on shared HPC systems. These delays are outside CIF’s control and must be considered when setting service expectations.

Security and compliance requirements in defence-oriented scenarios can also restrict network connectivity between exeuction sites, which may limit the exchange of data or the integration of human supervision tools. Our validation relied on secure transfer and authenticated endpoints, but larger deployments may require stronger safeguards, detailed logs of user actions for accountability, and integration with policy-aligned identity management.Finally, while CIF allows workflows to be changed at runtime, very large or uncontrolled modifications should be limited by safety rules to prevent errors or unstable states in long-running workflows.

\subsection{Summary}

In summary, existing frameworks perform well in their own domains such as enterprise automation, container-based workflows, or HPC scheduling, but they usually manage the human oversight as external or blocking, and they do not support runtime restructuring based on supervision. CIF combines configuration-based specification, asynchronous HITL, and hybrid HPC execution in a single workflow-oriented architecture. This makes it possible for workflows to adapt rapidly under operational conditions while keeping full use of computing resources. CIF can therefore serve as a practical foundation for mission-oriented workflows that require both adaptability and computational scale.

\section{Conclusions and Future Work}

This paper presented the CIF, a workflow-oriented architecture that uses the configuration file as the central element of execution. By declaring HITL checkpoints together as computational tasks, CIF turns supervision from a blocking step into an asynchronous mechanism that runs in parallel with large-scale computation. Validation on a tactical ship-detection case study showed that CIF maintains workflow execution across heterogeneous resources while enabling rapid, human-guided adaptation.

The results highlight two main contributions. First, CIF puts adaptability into practice through a configuration-based specification that unifies task dependencies, execution sites, and HITL checkpoints in a single declarative file. Second, CIF separates human review time from compute time, allowing supervisors to guide iterative retraining on HPC without leaving resources idle. Compared with existing workflow systems, CIF makes HITL an explicit, asynchronous feature while providing configuration-level control over hybrid execution.

The future work will extend CIF in different ways. As a short-term objective, we plan to demonstrate CIF across other domains, including disaster response and space situational awareness, to test its generality. Another interest is exploring interoperability with workflow standards (e.g. CWL, WDL, etc.), which would allow CIF to integrate more easily with existing ecosystems. As a mid-term objective, we plan to develop an API layer above the CIF-CLI that will allow users to integrate their own web applications as the user interface for interaction and visualization.

\section*{Acknowledgments}
This work was supported by the European Defence Fund (EDF) under Grant Agreement No. 101103770 (KOIOS Project). The authors acknowledge the resources provided by the Spanish Supercomputing Network (RES) at the Barcelona Supercomputing Center (BSC) on MareNostrum 5 [Activity ID: IM-2025-2-0033]. The authors also thank NTTDATA for their collaboration in defining and validating the tactical use case.

\bibliographystyle{spiebib} 
\bibliography{references}

\end{document}